**Reversible intrinsic-extrinsic crossover of the anomalous Hall effect in PrAlSi via Al/Si ratio tuning**

Huiyang Yang[1,2], Xiaohua Luo[1,2,3,*], Hongwei You[1,2], Wei Ren[1,2], Yuntao Gao[1,2], Changcai Chen[1,2], Zhishuo Zhang[1,2], Chunsheng Fang[1,2], and Shengcan Ma[1,2,3,*]

[1] *Jiangxi Provincial Key Laboratory of Magnetic Metallic Materials and Devices, College of Rare Earths, Jiangxi University of Science and Technology, Ganzhou 341000, China*

[2] *National Rare Earth Functional Materials Innovation Center, Jiangxi University of Science and Technology, Ganzhou 341000, China*

[3] *Jiangxi Provincial Center for Interdisciplinary Studies in Future-oriented Materials, Jiangxi University of Science and Technology, Ganzhou 341000, China.*

## Abstract

The anomalous Hall effect (AHE) in magnetic Weyl semimetals arises from distinct microscopic origins—intrinsic Berry curvature and extrinsic scattering mechanisms—yet reversible switching between these regimes within a single material platform remains elusive. Here, we demonstrate that off- stoichiometric tuning of the Al/Si ratio (from ~1.05 to ~0.75) reversibly drives an intrinsic–extrinsic–intrinsic crossover of the AHE in PrAlSi single crystals. Across four compositions spanning a narrow Si variation ($PrAl_{0.93}Si_{0.89}$ to $PrAl_{0.79}Si_{1.06}$), the anomalous Hall conductivity (AHC) exhibits a non-monotonic arch (170 → 1413 → 507 → 164 $\Omega^{-1}cm^{-1}$). The intermediate composition $PrAl_{0.90}Si_{0.94}$ (Al/Si ~0.96) exhibits high carrier mobility, multi-carrier transport, yielding a giant AHC of 1413 $\Omega^{-1}cm^{-1}$ dominated by extrinsic skew-scattering. The Si-poor and Si-rich compositions exhibit lower AHC values and intrinsic Berry-curvature-dominated AHE, establishing a composition-driven intrinsic–extrinsic–intrinsic crossover. Deviating from the intermediate composition suppresses mobility and reverses carrier type from hole-like to electron-like. The Al/Si ratio thus serves as a chemical knob that primarily modulates the Fermi level via compositional band filling, while off-stoichiometry-induced chemical disorder plays a subordinate role in determining the AHE mechanism. These results position off-stoichiometric PrAlSi as a tunable platform where subtle compositional variation reversibly switches the AHE mechanism.

*Corresponding authors: xhluo2018@jxust.edu.cn (Xiaohua Luo), mashengcan@jxust.edu.cn (Shengcan Ma)

# 1. Introduction

The anomalous Hall effect (AHE) in ferromagnetic metals comprises an intrinsic contribution arising from the Berry curvature of topological band structures and extrinsic contributions originating from disorder-mediated skew scattering and side jump [1-5]. The intrinsic anomalous Hall conductivity (AHC) is bounded by the integral of Berry curvature over occupied states and typically falls in the order of $10^2$–$10^3$ $\Omega^{-1}cm^{-1}$ for most magnetic topological materials [6-10]. In contrast, extrinsic skew scattering is not subject to this Berry-curvature limit and can produce a giant AHC that far exceeds the intrinsic value, often scaling as $\sigma_{\mathrm{AHE}} \propto \sigma^2$ in the moderately conductive regime [11-14]. A critical practical issue is that when skew scattering dominates, the colossal extrinsic signal completely masks the underlying intrinsic Berry contribution, making it impossible to assess the topological origin of the AHE from the magnitude alone.

The tetragonal rare-earth compounds RAlX (R = rare earth, X = Si, Ge) provide a representative platform for magnetic Weyl physics [15-17]. In the ordered LaPtSi-type structure, RAlX compounds crystallize in the noncentrosymmetric space group $I4_1md$ (No. 109), with R, Al, and X forming separate square-net layers along the *c* axis [Fig. 1(a)] [18]. The distinct 4*a* occupations break inversion symmetry, which stabilizes Weyl nodes already in the paramagnetic state [19,20]. Ferromagnetic order subsequently breaks time-reversal symmetry and produces type-I/type-II Weyl pairs, Fermi-arc surface states, topological and loop Hall effects, as well as anomalous Nernst responses and field- or temperature-induced band reconstructions [15–17,19-30]. The same family also shows a well-known structural complication: because Al and Si have similar scattering factors, site mixing can drive the system into the centrosymmetric α-$ThSi_2$-type $I4_1/amd$ (No. 141) with random 8*e* occupation [Fig. 1(b)], so that inversion breaking, Weyl topology, and disorder strength are sample dependent and must be specified rather than assumed [31].

Within this context, PrAlSi is a particularly instructive ferromagnetic member. Single crystals order ferromagnetically at $T_C \approx 17.8$ K with strong *c*-axis Ising anisotropy and a saturation moment of about 3 $\mu_B$/Pr [32,33]. Transport studies report semimetallic behavior, nonsaturating magnetoresistance, small Shubnikov–de Haas pockets, and a large anomalous Hall conductivity of order 2000 $\Omega^{-1}cm^{-1}$ [34]. More recent temperature-dependent ARPES finds that the low-energy Weyl-like bands change little across $T_C$, yet the extracted $\sigma_{AHE}$ can reach ~4200 $\Omega^{-1}cm^{-1}$ and follows $\sigma_{AHE} \propto \sigma^2$, demonstrating dominant extrinsic skew scattering rather than a simple Berry-curvature response [35]. Importantly, most published PrAlSi crystals are not strictly equiatomic. Self-flux samples have been reported with $I4_1/amd$ refinement and compositions such as $PrAl_{1.13}Si_{0.87}$, or with Si slightly richer than Al [29,33,34]. Al/Si disorder is known to affect residual resistivity and inversion symmetry, yet the systematic evolution of carrier density, Fermi-surface polarity, and AHE mechanism with controlled off-stoichiometric Al/Si variation has not been established. In these non-equiatomic crystals, the giant AHC far exceeds intrinsic Berry-curvature limits, extrinsic skew scattering dominates and overshadows the underlying Weyl/Berry contribution. The key challenge is therefore to systematically tune the Fermi level and disorder strength to suppress the extrinsic skew channel and unveil the intrinsic AHE.

In this work, we systematically vary the Al/Si ratio in PrAlSi single crystals to investigate how subtle compositional changes affect the AHE mechanism. Across four compositions spanning a narrow Si variation ($PrAl_{0.93}Si_{0.89}$ to $PrAl_{0.79}Si_{1.06}$), we observe a pronounced non-monotonic AHC evolution (170 → 1413 → 507 → 164 $\Omega^{-1}cm^{-1}$), accompanied by a systematic carrier-type reversal from hole-dominated (Si-poor) through multi-carrier (intermediate) to electron-dominated (Si-rich). The intermediate composition $PrAl_{0.90}Si_{0.94}$ lies at the extrinsic skew-scattering peak, where a giant AHC of 1413 $\Omega^{-1}cm^{-1}$ coincides with high carrier mobility and multi-carrier transport. By contrast, the Si-poor and Si-rich compositions exhibit intrinsic Berry-curvature-dominated AHE, establishing a reversible intrinsic–extrinsic–intrinsic crossover within a single material platform. These results imply that compositionally controlled Fermi-level shifting via Al/Si ratio, rather than disorder dose, is the dominant tuning parameter governing the AHE evolution across the PrAlSi series.

## 2. Experimental details

High-quality single crystals of off-stoichiometric PrAlSi were grown by the self-flux method using excess Al as the flux [30,34-36]. The starting materials were high-purity Pr ingots (99.9%), Al granules (99.999%), and Si pieces (99.999%). The raw materials were weighed according to the following nominal molar ratios: $PrAl_{10}Si_x$ ($x$ = 0.5, 1.0, 1.5, 2, and 4). The mixtures were loaded into alumina crucibles, sealed in evacuated quartz tubes under high vacuum, and heated to 1050 °C over 4 hours. After dwelling at 1050 °C for 12 h to ensure complete homogenization, the melts were slowly cooled to 750 °C at a rate of 2 °C/h. At 750 °C, the quartz tubes were removed from the furnace and the excess Al flux was removed by centrifugation. Shiny plate-like single crystals with typical dimensions of 2–5 mm were obtained.

Phase purity and crystal structure were examined by powder X-ray diffraction (XRD) using a PANalytical-Empyream diffractometer with Cu Kα radiation (λ = 1.5406 Å) at room temperature. The as-grown crystal surfaces were aligned parallel to the substrate plane, and diffraction patterns were collected over $2\theta$ = 10°–90° with a step size of 0.02°. Lattice parameters were refined by the Le Bail method using FullProf Suite. The actual chemical compositions of all as-grown crystals were determined by energy-dispersive X-ray spectroscopy (EDS, FEI MLA650F).

Magnetic and electrical transport properties were measured with the Quantum Design physical property measurement system (QD PPMS-9T). Temperature-dependent magnetization $M(T)$ was measured under both zero-field-cooled (ZFC) and field-cooled (FC) protocols in applied fields of 0.1 T. Isothermal magnetization $M(H)$ hysteresis loops were collected at various temperatures between 2 K and 40 K with the magnetic field applied parallel and perpendicular to the crystallographic *c* axis. Electrical transport measurements were carried out using the standard four-probe method. Electrical contacts were made using silver paste and 25 μm silver wires, with the current *I* parallel to the *ab* plane. Resistivity and Hall effect measurements were conducted using the same PPMS platform.

## 3. Results and discussion

Powder XRD patterns of the series crystals grown from nominal starting compositions $PrAl_{10}Si_x$ ($x$ = 0.5, 1.0, 1.5, 2, and 4) are shown in Fig. 1(c). As is common for RAlSi compounds, the noncentrosymmetric ordered model $I4_1md$ (LaPtSi-type) and the centrosymmetric disordered model $I4_1/amd$ (α-$ThSi_2$-type) yield nearly indistinguishable diffraction patterns when used to refine a single-phase sample due to the similar scattering factors of Al and Si [18,25]. In contrast to the $x \leq 2$ samples, the $x = 4$ sample exhibits several pronounced additional peaks (marked with asterisks) that can be unambiguously indexed to the Si-rich $I4_1/amd$ ($\alpha$-$ThSi_2$-type) structure. The remaining peaks in $x = 4$ overlap with those of the $x \leq 2$ samples and can be indexed to the noncentrosymmetric $I4_1md$ (LaPtSi-type) structure, consistent with second-harmonic-generation measurements that confirmed the noncentrosymmetric $I4_1md$ structure for PrAlSi [36]. This identifies $x = 4$ as a two-phase mixture of $I4_1md$ and $I4_1/amd$, rather than a single-phase disordered solid solution. The complete absence of $I4_1/amd$ peaks in the $x = 1$, 1.5 and 2 samples therefore confirms them as single-phase $I4_1md$. The $x = 0.5$ sample contains a trace amount of $Pr_3Al_{11}$ impurity (orthorhombic *Immm*) and Al flux, but its majority matrix is likewise assigned to $I4_1md$. Within the single-phase regime, the main $I4_1md$ peaks shift systematically to higher angles with increasing Si, indicating lattice contraction. Rietveld refinement profiles are provided in Fig. S1 and Table S1 of Supplemental Material [37]. Because Al and Si have similar scattering factors, site-occupancy refinement is unreliable; the extracted lattice parameters therefore serve only to track lattice evolution. Figure 1(d) shows the evolution of lattice parameters ($a$, $c$), $c/a$ ratio, and unit cell volume ($V$) of the LaPtSi-type phase as a function of nominal Si content $x$. The $c$-axis decreases continuously from $x = 0.5$ to $x = 2$, reflecting the successful substitution of smaller Si atoms for Al. In contrast, the $a$-axis exhibits negligible variation across the series, resulting in a systematic reduction of the $c/a$ ratio and unit cell volume $V$. Notably, the parameters exhibit only minimal further variation between $x = 2$ and $x = 4$, suggesting that Si incorporation into the LaPtSi-type lattice approaches its solubility limit near $x \approx 2$. Beyond this point, excess Si largely precipitates as a separate $ThSi_2$-type phase, as evidenced by the appearance of additional reflections at $x = 4$.

EDS results for the series crystals are summarized in Table S2 of Supplemental Material [37]. The actual compositions, normalized to Pr = 1, are $PrAl_{0.93}Si_{0.89}$ (A1), $PrAl_{0.90}Si_{0.94}$ (A2), $PrAl_{0.84}Si_{0.96}$ (A3), $PrAl_{0.79}Si_{1.06}$ (A4), $PrAl_{0.42}Si_{1.32}$ (A5) for crystals grown from nominal starting compositions $PrAl_{10}Si_x$ ($x = 0.5$, 1.0, 1.5, 2, and 4). For $x \leq 2$ (A1–A4), the Al content decreases systematically while the Si content increases, confirming successful growth of off-stoichiometric PrAlSi single crystals. The corresponding Al/Si ratios for A1–A5 are approximately 1.05, 0.96, 0.88, 0.75, and 0.32, respectively, confirming a continuous compositional shift from Al-rich to Si-rich. Notably, the A5 crystal isolated from the $x = 4$ batch exhibits extreme Al depletion (more than half that of A1–A4) and Si excess beyond 1.3, consistent with the $I4_1/amd$ phase identified by XRD. This crystal adopts a centrosymmetric structure fundamentally distinct from the noncentrosymmetric $I4_1md$ Weyl phase of A1–A4. Consequently, A5 is excluded from the further magnetic and transport analysis below, which focuses exclusively on the single-phase $I4_1md$ series (A1–A4).

Temperature-dependent magnetization curves $M(T)$ under 0.1 T with $H \,//\, c$ [Fig. 2(a)] show that all samples exhibit ferromagnetic transitions with $T_C$ ranging from 15.6 to 17.7 K (17.7, 16.4, 17.0 and 15.6 K for A1–A4). Isothermal magnetization curves at 2 K show saturation near 0.4 T along the $c$ axis for all compositions [Fig. 2(b)]. The saturation magnetization $M_S$ decreases from 3.0 $\mu_B$/f.u. for A1 to 2.6 $\mu_B$/f.u. for A4 as Si content increases. These $M_S$ values are consistent with theoretical $Pr^{3+}$ moments [33,34]. For $H$ // $ab$, $M$ increases nearly linearly without saturation up to 5 T, confirming strong magnetocrystalline anisotropy with the easy axis along $c$ [32,33]. The relatively narrow $T_C$ range and overall preservation of $Pr^{3+}$ moments across all four compositions indicate that varying the Al/Si ratio (from ~1.05 to ~0.75) induces only minor modifications to the magnetic properties of PrAlSi.

Figure 2(c) presents the temperature-dependent longitudinal resistivity $\rho_{xx}(T)$. All four samples exhibit metallic behavior with positive temperature coefficients. The residual resistivity ratio $RRR = \rho_{xx}(300\text{ K})/\rho_{xx}(2\text{ K})$ is 3.19, 3.81, 1.69, and 1.34 for A1-A4 samples, respectively. The relatively low $RRR$ values are common in this material family and have been attributed to chemical disorder in related Weyl semimetals [31]. Among them, the intermediate composition A2 shows the lowest residual resistivity ($\rho_{xx} \approx 22\ \mu\Omega \cdot \text{cm}$) and the highest $RRR$ (3.81), consistent with previously reported results in PrAlSi prepared from the same nominal ratio and identical growth process [34]. Magnetoresistance ($MR$) at 2 K [Fig. 2(d)] shows that A2 exhibits a large $MR$ (~110%) with a nearly linear field dependence, consistent with the previously established Weyl semimetal state in PrAlSi and the expected multi-carrier transport [29,34,35]. In contrast, deviating from the Al/Si ratio of A2 (~0.96) introduces additional chemical disorder. A3 and A4 display substantially higher residual resistivity and lower $RRR$ values than A2, indicating enhanced carrier scattering. Correspondingly, their magnetoresistance is lower by one to two orders of magnitude than A2, attributed to a combination of reduced carrier mobility and the breakdown of the multi-carrier transport regime, as confirmed by the Hall measurements below. Notably, A1 and A2 exhibit rather similar residual resistivity and *RRR* values, indicating that the slight Si increase from A1 to A2 does not introduce substantial extra disorder. However, A1 also shows a severe *MR* reduction, demonstrating that *MR* suppression in this system cannot be attributed simply to changes in scattering rate.

Figure 3(a) displays the Hall resistivity $\rho_{yx}(H)$ at 2 K and reveals distinct behaviors across the four compositions. The intermediate composition A2 exhibits a nonlinear Hall signal, indicative of multi-carrier transport, consistent with previous work [34,35]. This nonlinear response is physically supported by angle-resolved photoemission (ARPES) studies on PrAlSi, which map a multi-pocket Fermi surface with electron pockets near Γ and X and hole-like features along Γ–M [38]. In contrast, A1 shows a positive linear high-field slope, characteristic of hole-type carriers, whereas A3 and A4 show a negative linear high-field slope below $T_C$, confirming electron-type transport. The single-band behavior of A1, A3, and A4 suggests that Al/Si-ratio tuning shifts the Fermi level away from the multi-pocket crossing region observed in ARPES [37], consistent with the narrow energy scale (~25 meV) of band evolution across $T_C$ revealed by temperature-dependent ARPES [30]. Notably, A3 retains a negative slope above $T_C$, whereas A4 switches to a positive slope. This indicates that A4 approaches the electron-hole compensation regime, where the mobility balance evolves progressively with temperature. Together, these results demonstrate a continuous carrier-type reversal from hole-dominated to multi-carrier to electron-dominated, consistent with electron doping driven by the decreasing Al/Si ratio (from ~1.05 in A1 to ~0.75 in A4). Such evolution underscores the sensitivity of the transport regime to the Al/Si ratio.

To quantitatively decompose the Hall response, the nonlinear Hall data of A2 were analyzed using a two-band model combined with the anomalous contribution $R_S M_S$:

$$\rho_{\mathrm{H}} = \rho_{\mathrm{AHE}}(H) + \rho_{\mathrm{OHE}}(H) \quad (1)$$

$$\rho_{\mathrm{AHE}}(H) = R_S M_S \quad (2)$$

$$\rho_{\mathrm{OHE}} = \frac{(n_h \mu_h^2 - n_e \mu_e^2) + \mu_e^2 \mu_h^2 (n_h - n_e) H^2}{(n_e \mu_e + n_h \mu_h)^2 + \mu_e^2 \mu_h^2 (n_h - n_e)^2 H^2} \cdot \frac{H}{e} \quad (3)$$

Fitting yields $n_e \sim 3.3\times10^{20}$ cm$^{-3}$, $n_h \sim 0.9\times10^{19}$ cm$^{-3}$, $\mu_e \sim 742$ cm$^2$ V$^{-1}$ s$^{-1}$, and a remarkably high hole mobility $\mu_h \sim 4538$ cm$^2$ V$^{-1}$ s$^{-1}$ at 2 K [Fig. 3(d)]. To characterize the overall drift mobility of the two-carrier system A2, we adopt the conductivity-weighted average $\overline{\mu} = \sigma_{xx} / (en_e + en_h) = (n_e \mu_e + n_h \mu_h) / (n_e + n_h)$ from the two-band Drude model. This yields $\overline{\mu} \approx 843$ cm$^2$ V$^{-1}$ s$^{-1}$ at 2 K, which is close to $\mu_e$ because the electron term dominates $\sigma_{xx}$, although $\mu_h$ is remarkably high. The extracted AHC for A2 is 1413 $\Omega^{-1}\cdot$cm$^{-1}$ at 2 K [Fig. 3(b)], slightly smaller than the reported ~2000 $\Omega^{-1}$cm$^{-1}$ for PrAlSi [34]. The AHC of A2 (1413 $\Omega^{-1}\cdot$cm$^{-1}$) far exceeds the theoretical intrinsic value of ~400 $\Omega^{-1}\cdot$cm$^{-1}$ [35,36], consistent with extrinsic skew-scattering dominance. In contrast, linear fitting of the high-field Hall data using $\rho_{\mathrm{H}} = R_0 H + R_S M_S$ yields positive $R_0$ for A1, and negative $R_0$ for A3 and A4. Carrier concentration $n$ and mobility $\mu$ extracted from $|R_0| = 1/ne$ and $\sigma_{xx} = ne\mu$ at 2 K are $n_h \sim 1.5\times10^{21}$ cm$^{-3}$, $\mu_h \sim 137$ cm$^2$ V$^{-1}$ s$^{-1}$ for hole-dominated A1, $n_e \sim 3.2\times10^{21}$ cm$^{-3}$, $\mu_e \sim 25$ cm$^2$ V$^{-1}$ s$^{-1}$ for electron-dominated A3, and $n_e \sim 4.6\times10^{21}$ cm$^{-3}$, $\mu_e \sim 7.7$ cm$^2$ V$^{-1}$ s$^{-1}$ for electron-dominated A4 [Figs. 3(c), 3(e) and 3(f)]. The drastic mobility collapse in A3 and A4, accompanied by a substantial rise in carrier concentration, is attributed to off-stoichiometry-driven chemical disorder, including Al/Si anti-site defects and lattice disorder, which are corroborated by the elevated residual resistivity and the markedly suppressed magnetoresistance in both samples. A1 also shows a notable mobility reduction, despite its near-identical resistivity to A2, reflecting the loss of multi-carrier compensation rather than a drastic increase in scattering, consistent with the *MR* analysis. The corresponding AHC values are 170 $\Omega^{-1}\cdot cm^{-1}$ (A1), 507 $\Omega^{-1}\cdot cm^{-1}$ (A3) and 164 $\Omega^{-1}\cdot cm^{-1}$ (A4) at 2 K [Fig. 3(b)], respectively. These results underscore a disorder-sensitive transport regime: deviation from the Al/Si ratio of A2 introduces scattering centers that strongly suppress mobility. However, the AHC does not simply scale with mobility; A3 exhibits a much lower mobility than A1 (25 vs. 137 $cm^2\ V^{-1}\ s^{-1}$) yet a higher AHC (507 vs 170 $\Omega^{-1}\cdot cm^{-1}$), indicating that the extrinsic skew-scattering channel is strongly suppressed away from the A2 composition, while a finite intrinsic Berry-curvature contribution remains.

Notably, for each of A1, A3 and A4, the Karplus–Luttinger (KL) plot of $\rho_{yx}^{A}$ vs. $\rho_{xx}^{2}M$ at different temperatures collapses onto a respective straight line with a temperature-independent scaling coefficient $S_H$ [Figs. 4(a), 4(c) and 4(d)] [1,3]. Correspondingly, $\sigma_{xy}^{A}$ scales linearly with $M_S$ [Figs. 4(e), 4(g) and 4(h)]. Together with the AHC values being comparable to the theoretical intrinsic Berry-curvature limit (~400 $\Omega^{-1}\cdot cm^{-1}$), these results indicate that the AHE in these three compositions is intrinsically dominated by Berry curvature. By contrast, A2 fails to collapse in the KL plot [Fig. 4(b)] and exhibits noticeable departure of individual points from the linear fit of $\sigma_{xy}^{A}$–$M_S$ [Fig. 4(f)]; its inferred $S_H$ is not constant, consistent with extrinsic skew-scattering dominance. This composition-dependent crossover, from extrinsic skew-scattering dominance in the A2 composition to intrinsic Berry-curvature dominance in the off-stoichiometric A1, A3 and A4, highlights the delicate interplay between chemical composition and topological transport in PrAlSi.

The systematic evolution of transport and Hall properties across the off-stoichiometric PrAlSi series reveals that subtle Al/Si ratio tuning acts as an effective Fermi-level engineering tool in PrAlSi. Increasing the Si content continuously shifts the carrier type from hole-dominated (A1) to multi-carrier (A2) to electron-dominated (A3, A4), while the magnetic properties remain largely unchanged. This decoupling of carrier reversal from magnetic variation is consistent with Si-induced electron doping and an effective Fermi-level shift, as observed in the isostructural $PrAlGe_{1-x}Si_x$ series where increasing Si content progressively shrinks hole pockets and enlarges electron Fermi surfaces [36]. Notably, the endpoint PrAlGe remains intrinsically dominated, confirming that the extrinsic crossover in PrAlSi requires Si-driven Fermi-surface restructuring rather than merely inheriting the noncentrosymmetric Weyl topology [36]. The weak coupling between localized 4*f* moments and itinerant bands in PrAlSi, evidenced by ARPES across $T_C$ [38], further supports that the carrier-type evolution is driven by chemical tuning of the Fermi level rather than magnetism-induced band reconstruction. The Fermi-level shift—and the associated compositional band filling—drives a pronounced non-monotonic AHE evolution: the AHC rises from 170 $\Omega^{-1}cm^{-1}$ (A1) to a giant peak of 1413 $\Omega^{-1}cm^{-1}$ (A2), then falls back to 507 $\Omega^{-1}cm^{-1}$ (A3) and 164 $\Omega^{-1}cm^{-1}$ (A4) (Fig. 5). Within the unified quantum transport theory of the anomalous

Hall effect in ferromagnets [39], this arch-like behavior reflects a continuous intrinsic–extrinsic–intrinsic crossover: as the Fermi level traverses the multi-carrier window, extrinsic skew scattering is amplified in the super-clean regime, whereas single-carrier dominant regimes at both ends suppress this channel and restore intrinsic Berry-curvature dominance. In A1, A3, and A4, the AHC is governed primarily by intrinsic Berry curvature. The values of AHC for A1 and A4 fall within the theoretical intrinsic limit (~400 $\Omega^{-1}$ $cm^{-1}$ [36]), while that of A3 modestly exceeds it (Fig. 5), which may be attributed to its Fermi level lying closer to a secondary Weyl pocket [30,36] or to a partially retained extrinsic skew-scattering tail near the edge of the multi-carrier window. This supports the interpretation that compositionally controlled Fermi-level shifting continuously tunes the intrinsic Berry-curvature contribution. The high sensitivity of the AHC to the Al/Si ratio, encompassing carrier polarity reversal, is a characteristic signature pointing to Fermi-level tuning across a topological transport regime.

Prior efforts to tune the AHE in topological and Weyl magnets have predominantly relied on two strategies: heterogeneous element doping and hydrostatic pressure. Within the RAlX family, continuous Ge/Si substitution in $PrAlGe_{1-x}Si_x$ drives a largely unidirectional intrinsic-to-extrinsic crossover as the Fermi level shifts and extrinsic scattering accumulates [36]. A complementary case is $CeAlSi_{1-x}Ge_x$, where continuous Ge enrichment reverses carrier type from electron- to hole-like near $x \approx 0.5$–0.6, and progressively suppresses ferromagnetism toward incommensurate/antiferromagnetic order with enhanced chiral Weyl interactions; however, the AHE evolution is strongly entangled with magnetic-texture changes rather than a clean band-filling scan [40]. Outside RAlX, hetero-doping in kagome $Co_3Sn_2S_2$ primarily enhances extrinsic skew scattering (e.g., Fe/Ni substitution raises the AHE angle without restoring intrinsic Berry dominance) [41-43]. In parallel, hydrostatic pressure offers an impurity-free Fermi-level knob, most extensively in CeAlSi, where compression moves Weyl nodes and produces an AHE sign change near 5.6 GPa—but such pathways are often unidirectional and truncated by phase boundaries [27]. In contrast, off-stoichiometric Al/Si tuning in PrAlSi achieves a reversible intrinsic–extrinsic–intrinsic AHE arch within a single noncentrosymmetric lattice, distinct from the unidirectional crossover of hetero-doping and the phase-limited pressure routes.

Nevertheless, off-stoichiometric tuning of the Al/Si ratio inevitably introduces chemical disorder such as Al/Si anti-site defects and lattice defects, consistent with the progressively suppressed carrier mobility in A3/A4 and the markedly reduced magnetoresistance across the series. Mobility suppression alone cannot fully account for the AHE evolution: while A2→A4 shows a parallel decline in both mobility and AHC, A3 exhibits a lower mobility than A1 yet a higher AHC (Fig. 5). Within the conventional skew-scattering picture, the extrinsic skew contribution to the AHC scales as $\sigma_{xy}^{\mathrm{skew}} \sim u_{\mathrm{imp}} E_{\mathrm{SOC}} D \tau$ with $\tau \propto \mu$, where $u_{\mathrm{imp}}$ is the impurity potential, $E_{\mathrm{SOC}}$ is the spin–orbit coupling energy, $D$ is the density of states at the Fermi level, and $\tau$ is the relaxation time [39,44]. Because $\tau$ decreases with increasing scattering centers, a skew-scattering scenario would predict a lower AHC in A3 than in A1, contradicting the observed A3 > A1 trend (Fig. 5). This does not rule out an enhanced extrinsic skew channel at A2, but it indicates that the A1→A3 enhancement is not governed by disorder dose alone. Moreover, intrinsic AHC is nearly mobility-insensitive [45,46]. The present AHC–mobility decoupling therefore indicates that Al/Si-ratio-driven Fermi-level shift and compositional band filling—which primarily modulate *D* and the Berry-curvature-weighted bands rather than acting as a mere impurity-dose knob—are the dominant factors governing the AHE crossover. Future ARPES measurements on these off-stoichiometric single crystals would directly map the Fermi-surface evolution across the Al/Si series and provide microscopic confirmation of the proposed Fermi-level shift and multi-carrier window.

## 4. Conclusions

In summary, we have shown that subtle off-stoichiometric variations of the Al/Si ratio in PrAlSi single crystals enable reversible switching of the AHE between extrinsic and intrinsic regimes within a single noncentrosymmetric lattice. This compositional tuning acts as an effective Fermi-level engineering tool, driving a continuous carrier-type reversal and a non-monotonic AHC evolution that forms an intrinsic–extrinsic–intrinsic arch. The intermediate composition $PrAl_{0.90}Si_{0.94}$ exhibits a giant extrinsic skew-scattering peak, while compositions away from it retain intrinsic Berry-curvature-dominated responses. The observed decoupling of the AHC from carrier mobility demonstrates that compositional band filling, rather than disorder-induced scattering alone, governs the crossover. These results position off-stoichiometric PrAlSi as a tunable platform for systematically exploring the interplay of composition and topology, with potential extension to other members of the RAlX family.

**Acknowledgment**

This work has been supported by the National Natural Science Foundation of China (Grant Nos. 52561034, 52561035 and 52101223), the Natural Science Foundation of Jiangxi Province (Grant No. 20242BAB25212), and Jiangxi Provincial Key Laboratory of Magnetic Metallic Materials and Devices (Grant No. 2024SSY05061). Meanwhile, the authors express a sincere gratitude to the Pioneer Research Program of Jiangxi Provincial Center for Interdisciplinary Studies in Future-oriented Materials for the financial support.

# Reference

[1] N. Nagaosa, J. Sinova, S. Onoda, A. H. MacDonald, and N. P. Ong, Anomalous Hall effect, Rev. Mod. Phys. 82, 1539 (2010).

[2] S. Onoda, N. Sugimoto, and N. Nagaosa, Intrinsic versus extrinsic anomalous Hall effect in ferromagnets, Phys. Rev. Lett. 97, 126602 (2006).

[3] R. Karplus and J. M. Luttinger, Hall effect in ferromagnetics, Phys. Rev. 95, 1154 (1954).

[4] J. Smit, Spontaneous Hall effect in ferromagnetics, Physica 24, 39 (1958).

[5] L. Berger, Side-jump mechanism for the Hall effect of ferromagnets, Phys. Rev. B 2, 4559 (1970).

[6] D. Xiao, M. C. Chang, and Q. Niu, Berry-phase effects on electronic properties, Rev. Mod. Phys. 82, 1959 (2010).

[7] S. Nakatsuji, N. Kiyohara, and T. Higo, Large anomalous Hall effect in noncollinear antiferromagnet $Mn_3Sn$, Nature 527, 212 (2015).

[8] E. Liu, Y. Sun, N. Kumar, L. Muechler, A. Sun, L. Jiao, S.-Y. Yang, D. Liu, A. Liang, Q. Xu, et al., Giant anomalous Hall effect in a ferromagnetic kagome-lattice semimetal, Nat. Phys. 14, 1125 (2018).

[9] A. Sakai, Y. P. Mizuta, A. A. K. Nugroho, R. Sihombing, T. Koretsune, M.-T. Suzuki, N. Takemori, R. Ishii, D. Nishio-Hamane, R. Arita, et al., Giant anomalous Hall effect in the chiral itinerant ferromagnet $Co_2MnGa$, Nat. Phys. 14, 1119 (2018).

[10] W. Ma, X. Xu, J.-X. Yin, H. Yang, H. Zhou, Z.-J. Cheng, Y. Huang, Z. Qu, F. Wang, M. Z. Hasan, and S. Jia, Rare Earth Engineering in $RMn_6Sn_6$ (R = Gd–Tm, Lu)

Topological Kagome Magnets, Phys. Rev. Lett. 126, 246602 (2021).

[11] Y. Shiomi, Y. Onose, and Y. Tokura, Extrinsic AHE in pure iron and doped alloys, Phys. Rev. B 79, 100404(R) (2009).

[12] J. Chen, X. Yang, F. Zhou, Y.-C. Lau, W. Feng, Y. Yao, Y. Li, Y. Jiang, and W. Wang, Colossal anomalous Hall effect in the layered antiferromagnetic $EuAl_2Si_2$ compound, Mater. Horiz. 11, 4665–4673 (2024).

[13] S. Xu, J.-F. Zhang, S.-X. Li, J. Lin, X. Ma, W. Yang, J.-J. Mi, Z. Li, T.-H. Li, Y.-Y. Wu, J. Ma, Q. Tao, W.-H. Jiao, X. Xu, Z. Zhu, Y. Xu, H. Guo, T.-L. Xia, and Z.-A. Xu, Giant anomalous Hall conductivity in frustrated magnet $EuCo_2Al_9$, Mater. Today 95, 103285 (2026).

[14] S.-Y. Yang, Y. Wang, B. R. Ortiz, D. Liu, J. Gayles, E. Derunova, R. González-Hernández, L. Šmejkal, Y. Chen, S. S. P. Parkin, S. D. Wilson, E. S. Toberer, T. McQueen, and M. N. Ali, Giant unconventional anomalous Hall effect in the metallic frustrated magnet candidate $KV_3Sb_5$, Sci. Adv. 6, eabb6003 (2020).

[15] G. Chang, B. Singh, S.-Y. Xu, D.S. Sanchez, N. Alidoust, M.M. Deniz, H. Zheng, C. Su, G. Bian, H. Lin, and M.Z. Hasan, Magnetic and noncentrosymmetric Weyl fermion semimetals in the RAlGe family of compounds (R = rare earth), Phys. Rev. B 97, 041104(R) (2018).

[16] M.S. Alam, A. Alfonsov, M. Binder, W. Schnelle, R. Hull, G.H. Fecher, C. Felser, J. van den Brink, B. Büchner, and S. Wurmehl, Sign change of anomalous Hall effect and anomalous Nernst effect in the Weyl semimetal CeAlSi, Phys. Rev. B 107, 085102 (2023).

[17] J. Gaudet, C. Dalpra, M. B. Stone, J. A. Rodriguez-Rivera, D. R. Yahne, K. Ross, J. S. White, R. Sibille, M. C. Rahn, N. R. Hayre, R. Coldea, L. D. Sanjeewa, M. A. McGuire, D. Mandrus, G. M. Luke, and B. D. Gaulin, Weyl-mediated helical magnetism in NdAlSi, Nat. Mater. 20, 1650 (2021).

[18] M. A. Avila, Crystal chemistry of RAlSi and related phases, J. Alloys Compd. 317–318, 273 (2001).

[19] D. S. Sanchez, G. Chang, I. Belopolski, H. Lu, J.-X. Yin, N. Alidoust, X. Xu, T. A. Cochran, X. Zhang, Y. Bian, et al., Observation of Weyl fermions in a magnetic non-centrosymmetric crystal, Nat. Commun. 11, 3356 (2020).

[20] C. Li, J. Zhang, Y. Wang, H. Liu, Q. Guo, E. Rienks, W. Chen, F. Bertran, H. Yang, D. Phuyal, et al., Emergence of Weyl fermions by ferrimagnetism in a noncentrosymmetric magnetic Weyl semimetal, Nat. Commun. 14, 7185 (2023).

[21] T. Suzuki, L. Savary, J.-P. Liu, J. W. Lynn, L. Balents, and J. G. Checkelsky, Singular angular magnetoresistance in a magnetic nodal semimetal, Science 365, 377 (2019).

[22] H.-Y. Yang, B. Singh, J. Gaudet, B. Lu, C.-Y. Huang, W.-C. Chiu, S.-M. Huang, B. Wang, F. Bahrami, B. Xu, J. Franklin, I. Sochnikov, D. E. Graf, G. Xu, Y. Zhao, C. M. Hoffman, H. Lin, D. H. Torchinsky, C. L. Broholm, A. Bansil, and F. Tafti, Noncollinear ferromagnetic Weyl semimetal with anisotropic anomalous Hall effect, Phys. Rev. B 103, 115143 (2021).

[23] P. Puphal, V. Pomjakushin, N. Kanazawa, Y. Nambu, J. S. White, S. H. Lee, T. Onishi, K. Nakajima, M. Kriener, Y. Taguchi, et al., Topological magnetic phase in

the candidate Weyl semimetal CeAlGe, Phys. Rev. Lett. 124, 017202 (2020).

[24] D. Destraz, L. Das, S. S. Tsirkin, Y. Xu, T. Neupert, J. Chang, A. Schilling, A. G. Grushin, J. Kohlbrecher, L. Keller, et al., Magnetism and anomalous transport in the Weyl semimetal PrAlGe: possible route to axial gauge fields, npj Quantum Mater. 5, 5 (2020).

[25] B. Meng, H. Wu, Y. Qiu, C. Wang, Y. Liu, Z. Xia, S. Yuan, H. Chang, and Z. Tian, Large anomalous Hall effect in the ferromagnetic Weyl semimetal candidate PrAlGe, APL Mater. 7, 051110 (2019).

[26] E. Cheng, L. Yan, X. Shi, R. Lou, A. Fedorov, M. Behnami, Tunable positions of Weyl nodes via magnetism and pressure in the ferromagnetic Weyl semimetal CeAlSi, Nat. Commun. 15, 1467 (2024).

[27] Y. Sun, C. Lee, H.-Y. Yang, D. H. Torchinsky, F. Tafti, and J. Orenstein, Mapping domain-wall topology in the magnetic Weyl semimetal CeAlSi, Phys. Rev. B 104, 235119 (2021).

[28] N. Zhang, D. Wang, X. Zhang, S. Gao, H. Zhou, Y. Zhao, Z. Shi, P. Yang, Z. Yang, and Y. Zhang, Abnormally enhanced Hall Lorenz number in the magnetic Weyl semimetal NdAlSi, Nat. Commun. 15, 10255 (2024).

[29] L. Wu, S. Chi, H. Zuo, G. Xu, L. Zhao, Y. Luo, and Z. Zhu, Field-induced Lifshitz transition in the magnetic Weyl semimetal candidate PrAlSi, npj Quantum Mater. 8, 4 (2023).

[30] K. Wang, W. Shi, W. Cao, X. Yang, Z. Lv, C. Peng, C. Chen, D. Liu, H. Yang, L. Yang, M. Lyu, P. Sun, E. Liu, M. Ye, Y. Chen, Y. Sun, Y. Qi, and Z. Liu, Weyl Fermion Manipulation Through Magnetic Transitions in the Ferromagnetic Non-Centrosymmetric Weyl Semimetal PrAlSi, Adv. Electron. Mater. 11, 2500044

(2025).

[31] T. Wang, Y. Guo, and C. Wang, Structure and magnetic properties of RAlSi (R = light rare earth), Chin. Phys. B 30, 075102 (2021).

[32] L. Yang, Y. Sun, X. Deng, W. Cao, X. Ma, Y. Xiao, Z. Wang, Z. Hu, X. Hao, Y. Yuan, Z. Qin, W. Luo, Q. Ren, X. Tong, M. Aouane, M.D. Le, Y. Shi, Y. Qi, D. Adroja, and H. Luo, Crystalline electric field excitations in the Weyl semimetals RAlSi (R = Ce, Pr, Nd), Phys. Rev. B 112, 054439 (2025).

[33] M. Lyu, Z. Wang, K. R. Kumar, H. Zhao, J. Xiang, and P. Sun, Large anisotropic magnetocaloric effect in ferromagnetic semimetal PrAlSi, J. Appl. Phys. 127, 193903 (2020).

[34] M. Lyu, H. Bai, Y. Lu, J. Miao, Y. Wu, X. Yang, C. Xi, Y. Yang, D. Meng, X. Wang, Y. Zhang, T. Hu, and J. Wang, Nonsaturating magnetoresistance, anomalous Hall effect, and magnetic quantum oscillations in the ferromagnetic semimetal PrAlSi, Phys. Rev. B 102, 085143 (2020).

[35] H. Rong, A. Zhang, F. Yu, M. Lyu, H. Li, M. Arita, Z. Hao, Q. Jiang, Z. Liu, M. Ye, Y. Guo, J. Xu, C. Liu, J. Wang, S. Guo, L. Wang, J. Lin, K. Shimada, E. Liu, J. He, Q. Liu, and C. Chen, Dichotomy of magnetic effect between Weyl fermions and anomalous Hall effect in PrAlSi, Commun. Mater. 6, 90 (2025).

[36] H.-Y. Yang, B. Singh, B. Lu, C.-Y. Huang, F. Bahrami, W.-C. Chiu, D. Graf, S.-M. Huang, B. Wang, H. Lin, D. Torchinsky, A. Bansil, and F. Tafti, Transition from intrinsic to extrinsic anomalous Hall effect in the ferromagnetic Weyl semimetal $PrAlGe_{1-x}Si_x$, APL Mater. 8, 011111 (2020).

[37] See Supplemental Material for more information on the structural, composition magnetization, and Hall analysis of the samples.

[38] R. Lou, A. Fedorov, L. Zhao, A. Yaresko, B. Büchner, and S. Borisenko, Signature of weakly coupled *f* electrons and conduction electrons in magnetic Weyl semimetal candidates PrAlSi and SmAlSi, Phys. Rev. B 107, 035158 (2023).

[39] S. Onoda, N. Sugimoto, and N. Nagaosa, Quantum transport theory of anomalous electric, thermoelectric, and thermal Hall effects in ferromagnets, Phys. Rev. B 77, 165103 (2008).

[40] X. Pang, G. Peng, H. Zhang, B. Qian, X. Jiang, Z. Tian, and Y. Yang, Investigation on crystal structure, electrical and magnetic properties of $CeAl(Si_{1-x}Ge_x)$, J. Solid State Chem. 308, 122877 (2022).

[41] J. Shen, Q. Zeng, S. Zhang, H. Sun, Q. Yao, X. Xi, W. Wang, G. Wu, B. Shen, Q. Liu, and E. Liu, 33% Giant Anomalous Hall Current Driven by Both Intrinsic and Extrinsic Contributions in Magnetic Weyl Semimetal $Co_3Sn_2S_2$, Adv. Funct. Mater. 30, 2000830 (2020).

[42] B. He, M. Yao, Y. Pan, K. E. Arpino, D. Chen, F. M. Serrano-Sánchez, S. Ju, M. Sun, Y. Sun, and C. Felser, Enhanced Weyl semimetal signature in $Co_3Sn_2S_2$ kagome ferromagnet by chlorine doping, Commun. Mater. 5, 275 (2024).

[43] L. Cao, I. A. Malik, Y. Wu, S.-T. Sun, Y.-Y. Lv, S.-H. Yao, J. Zhou, Z. Xu, C. Chen, V. L. Zhang, T. Yu, Y. Chen, D. Wu, and Y.-F. Chen, Significantly unconventional enhancement of anomalous Hall angle in magnetic Weyl semimetal $Co_3Sn_2S_{2-x}Se_x$, Adv. Mater. 38, e22443 (2026).

[44] Y. Feng, M. Lyu, J. Yang, S. Zhang, Y. Wang, Q. Ren, B. Wang, H. Wei, B. Shen, and E. Liu, Intrinsic and extrinsic dual-component evolution of electronic transport in a magnetic Weyl semimetal, Phys. Rev. B 114, 045131 (2026).

[45] L. Ding, J. Koo, L. Xu, X. Li, X. Lu, L. Zhao, Q. Wang, Q. Yin, H. Lei, B. Yan,

et al., Intrinsic anomalous Nernst effect amplified by disorder in a half-metallic semimetal, Phys. Rev. X 9, 041061 (2019).

[46] J. Liu, L. Ding, L. Xu, X. Li, K. Behnia, and Z. Zhu, Tuning the anomalous Nernst and Hall effects with shifting the chemical potential in Fe-doped and Ni-doped $Co_3Sn_2S_2$, J. Phys.: Condens. Matter 35, 375501 (2023).

**Figure caption**

**FIG. 1.** Crystal structure of (a) the LaPtSi-type phase (space group $I4_1md$) and (b) $ThSi_2$-type phase (space group $I4_1/amd$). Pr, Al, and Si atoms are shown as brown, gray, and green spheres, respectively. (c) Powder XRD patterns (left) and an enlarged view of the $2\theta$ = 32–33° region (right) for A1–A5 samples. Diamond symbols denote additional reflections from the Si-rich $ThSi_2$-type secondary phase at A5. (d) Evolution of lattice parameters ($a$, $c$), $c/a$ ratio, and unit cell volume ($V$) of the LaPtSi-type phase as a function of nominal Si content $x$.

**FIG. 2.** (a) Temperature-dependent magnetization $M(T)$ under a 0.1 T field applied along the $c$ axis ($H$ // $c$) for $PrAl_{0.93}Si_{0.89}$ (A1), $PrAl_{0.90}Si_{0.94}$ (A2), $PrAl_{0.84}Si_{0.96}$ (A3), and $PrAl_{0.79}Si_{1.06}$ (A4) samples. (b) Isothermal magnetization $M(H)$ loops at 2 K with the field applied along the $c$ axis ($H$ // $c$) and within the $ab$ plane ($H$ // $ab$). (c) Temperature dependence of longitudinal resistivity $\rho_{xx}(T)$ measured under zero field with current flowing in the $ab$ plane ($I$ // $ab$). (d) Magnetoresistance $MR$ at 2 K with the field applied along the $c$ axis.

**FIG. 3.** (a) Magnetic-field-dependent Hall resistivity $\rho_{yx}(H)$ measured at 2K with $I$ // $ab$ and $H$ // $c$. (b) Temperature dependence of the extracted anomalous Hall conductivity $\sigma_{xy}^{A}$. (c)–(f) Temperature dependence of carrier density $n$ (left axis) and mobility $\mu$ (right axis) for A1, A2, A3, and A4, respectively.

**FIG. 4.** (a)-(d) Anomalous Hall resistivity $\rho_{yx}^{A}$ plotted against $\rho_{xx}^{2}M$ at various

temperatures (2–16 K) for A1, A2, A3, and A4, respectively. Insets: Temperature dependence of the scaling coefficient $S_H$. (e)–(h) Anomalous Hall conductivity $\sigma_{xy}^{A}$ plotted against saturation magnetization $M_S$ with linear fitting (red lines).

**FIG. 5.** Comparison of anomalous Hall conductivity $\sigma_{xy}^{A}$ and mobility $\mu$ across the A1–A4 series at 2 K. The conductivity-weighted average mobility is adopted for A2. The red dashed line indicates the theoretical intrinsic Berry curvature limit (~400 $\Omega^{-1}$ $cm^{-1}$). The top shaded regions classify the dominant mechanism as intrinsic-dominated (A1), extrinsic skew-scattering (A2), and intrinsic-dominated (A3 and A4) from left to right.

**FIG. 1**

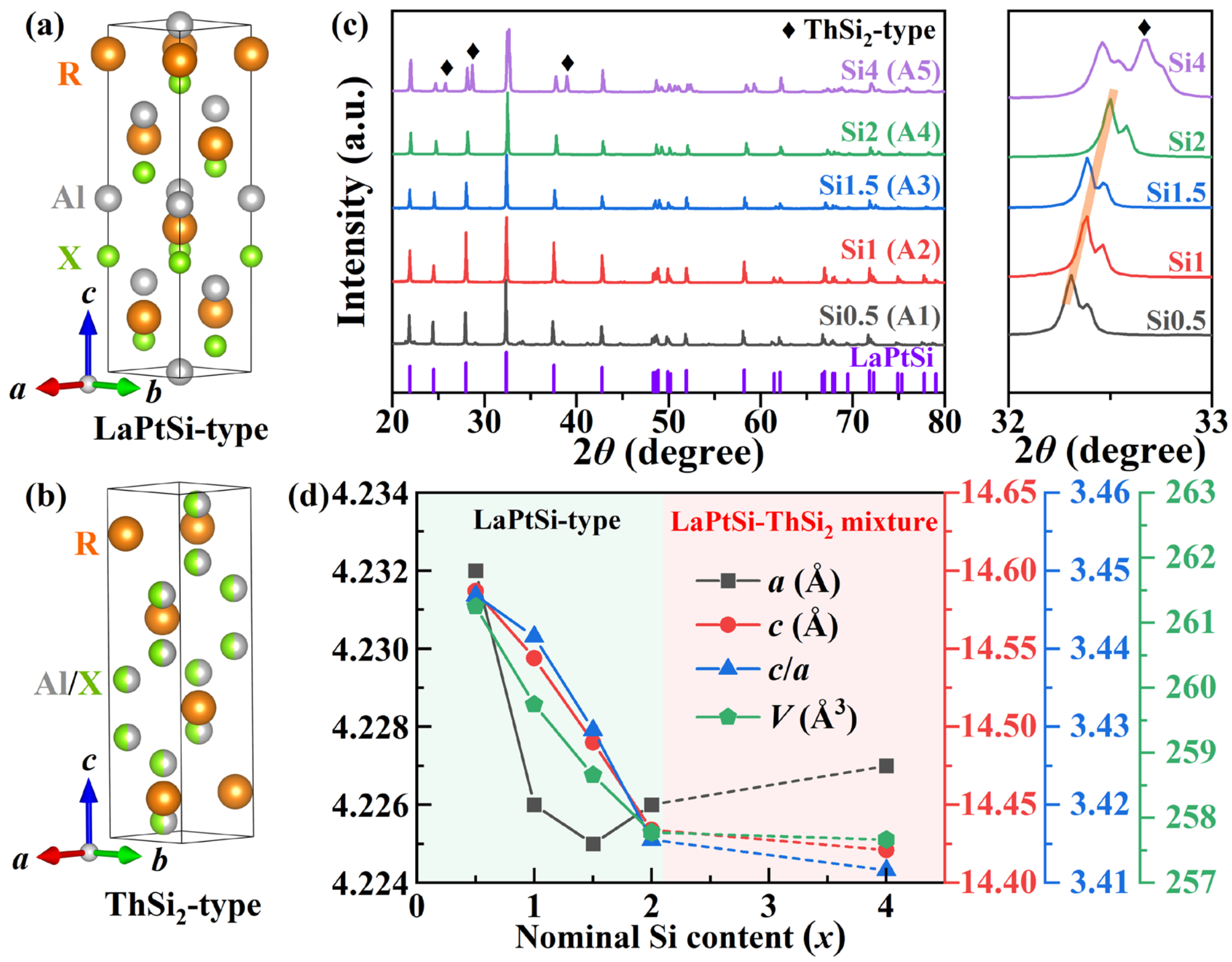

(a)
R
Al
X
c
a
b
LaPtSi-type
(b)
R
Al/X
c
a
b
ThSi2-type
(c)
♦ ThSi2-type
Si4 (A5)
Si2 (A4)
Si1.5 (A3)
Si1 (A2)
Si0.5 (A1)
LaPtSi
Intensity (a.u.)
20 30 40 50 60 70 80
2θ (degree)
Si4
Si2
Si1.5
Si1
Si0.5
32 33
2θ (degree)
(d)
LaPtSi-type
LaPtSi-ThSi2 mixture
a (Å)
c (Å)
c/a
V (Å3)
4.224 4.226 4.228 4.230 4.232 4.234
14.40 14.45 14.50 14.55 14.60 14.65
3.41 3.42 3.43 3.44 3.45 3.46
257 258 259 260 261 262 263
0 1 2 3 4
Nominal Si content (x)

**FIG. 2**

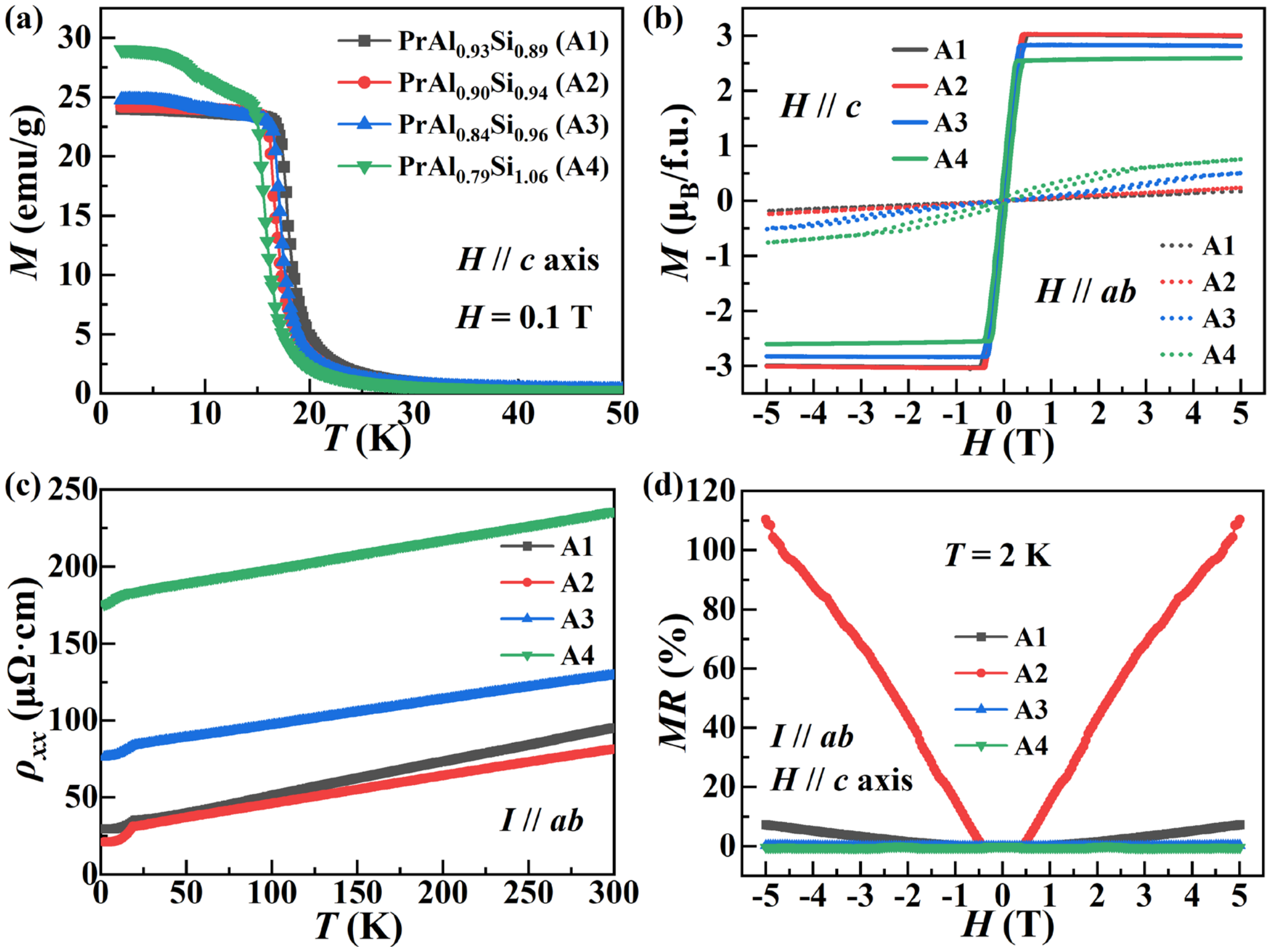

(a)
PrAl0.93Si0.89 (A1)
PrAl0.90Si0.94 (A2)
PrAl0.84Si0.96 (A3)
PrAl0.79Si1.06 (A4)
H // c axis
H = 0.1 T
M (emu/g)
T (K)
(b)
H // c
A1
A2
A3
A4
H // ab
M (μB/f.u.)
H (T)
(c)
I // ab
ρxx (μΩ·cm)
T (K)
(d)
T = 2 K
I // ab
H // c axis
MR (%)
H (T)

FIG. 3

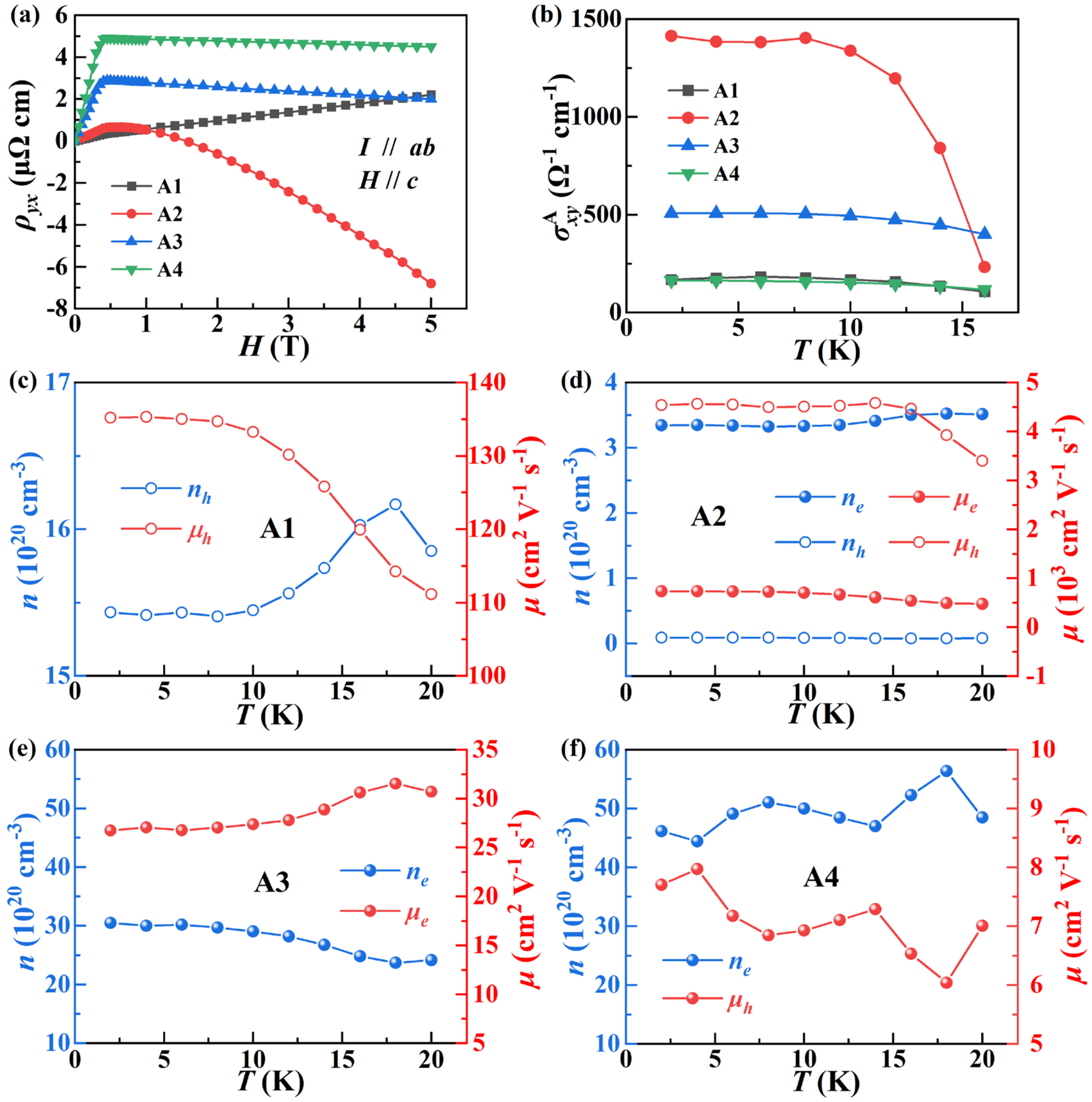

## FIG. 4

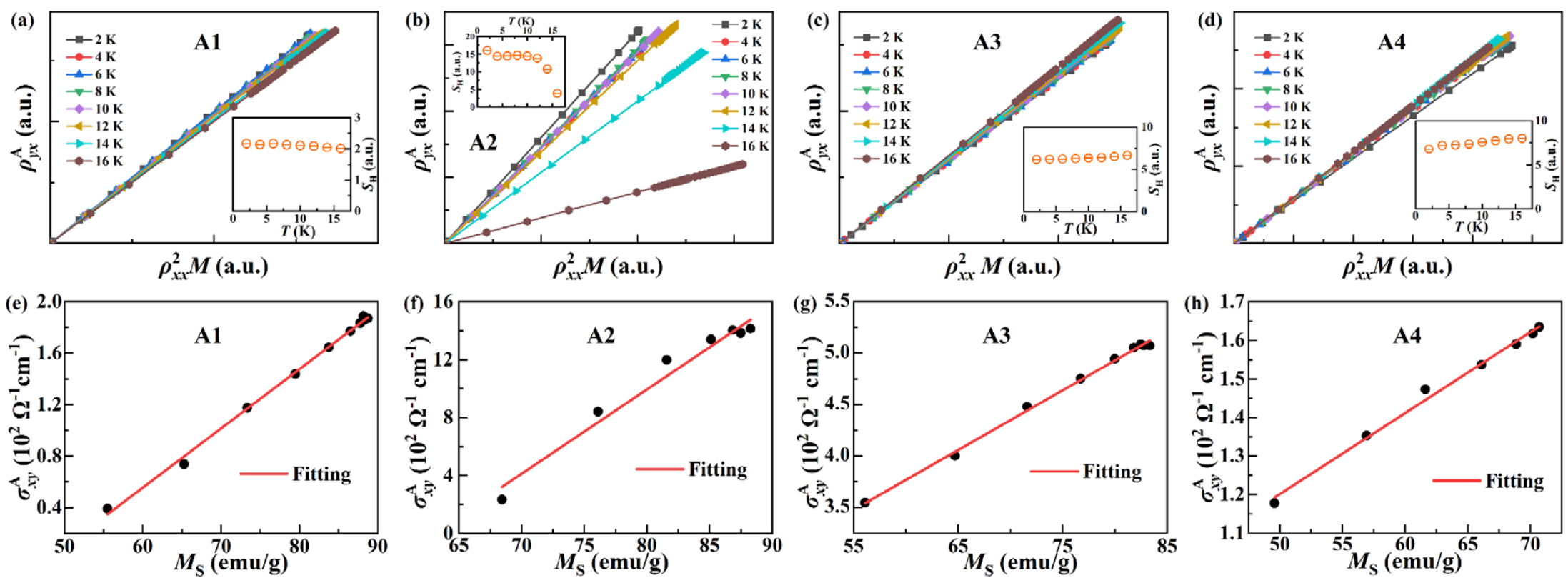

(a)
A1
2 K
4 K
6 K
8 K
10 K
12 K
14 K
16 K
$\rho_{yx}^{A}$ (a.u.)
$\rho_{xx}^{2}M$ (a.u.)
$T$ (K)
$S_H$ (a.u.)
(b)
A2
(c)
A3
(d)
A4
(e)
$\sigma_{xy}^{A}$ ($10^2\ \Omega^{-1}\mathrm{cm}^{-1}$)
$M_S$ (emu/g)
Fitting
(f)
(g)
(h)

**FIG. 5**

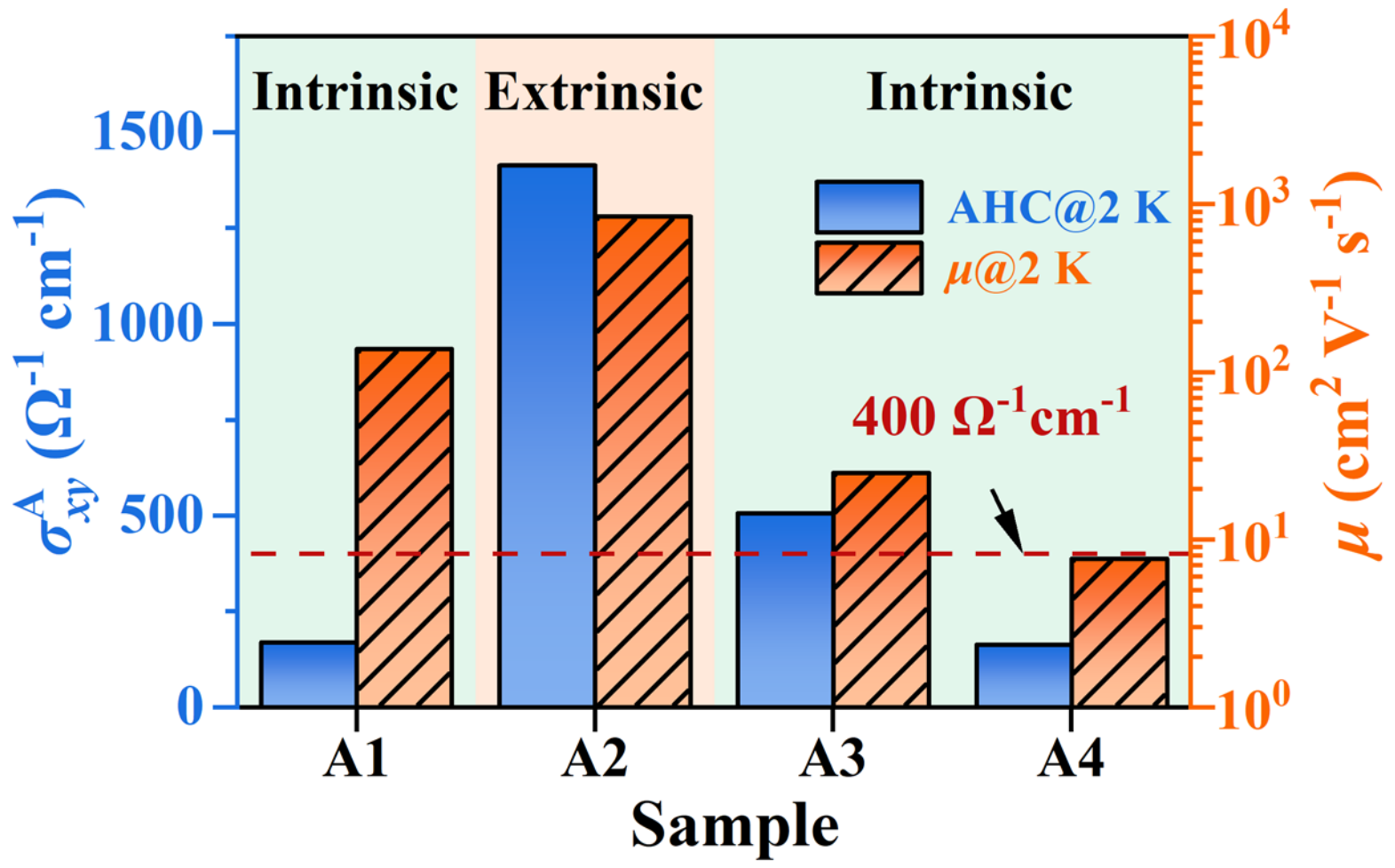

Intrinsic
Extrinsic
Intrinsic
AHC@2 K
μ@2 K
400 Ω-1cm-1
1500
1000
500
0
$\sigma_{xy}^{A}$ (Ω-1 cm-1)
10^4
10^3
10^2
10^1
10^0
μ (cm2 V-1 s-1)
A1
A2
A3
A4
Sample